\documentclass[]{mn2e}
\usepackage{times}
\usepackage{epsfig}
\usepackage{amssymb}
\usepackage{amsmath} 
\title[The H-deficient knot of Abell 58]
{The hydrogen-deficient knot of the `born again' planetary nebula Abell 58 (V605 Aql)}
\author[R. Wesson et al.]
{R. Wesson$^1$,  M.J. Barlow$^1$, X-W. Liu$^2$, P.J. Storey$^1$, B. Ercolano$^3$, O. De Marco$^4$\\
$^1$Department of Physics and Astronomy, University College London, Gower Street, London WC1E 6BT, UK\\
$^2$Department of Astronomy, Peking University, Beijing 100871, P. R. China\\
$^3$Harvard Smithsonian Center for Astrophysics, 60 Garden Street, Cambridge, MA 02138, USA\\
$^4$Department of Astrophysics American Museum of Natural History, Central Park West at 79th Street, NY 10024, USA\\}
\date{Received:}

\begin{document}
\maketitle

\begin{abstract}

We have analysed deep optical spectra of the `born-again' planetary nebula Abell~58 and its hydrogen-deficient knot, surrounding V605~Aql, which underwent a nova-like eruption in 1919.  Our analysis shows that the extinction towards the central knot is much higher than previously thought, with $c({\rm H}\beta)$=2.0.  The outer nebula is less reddened, with $c({\rm H}\beta)$=1.04.  We find that the outer nebula has a Ne/O ratio higher than the average PN value.

The electron temperature we derive for the central knot varies widely depending on the diagnostic used.  The [O~{\sc iii}] nebular-to-auroral transition ratio gives a temperature of 20\,800\,K, while the ratio of the [N~{\sc ii}] nebular and auroral lines give T$_{\rm e}$=15\,200\,K.  The helium line ratios $\lambda$5876/$\lambda$4471 and $\lambda$6678/$\lambda$4471 imply temperatures of 350\,K and 550\,K respectively.  Weakly temperature-sensitive O~{\sc ii} recombination line ratios imply similarly low electron temperatures.  Abundances derived from recombination lines are vastly higher than those found from collisionally excited lines, with the abundance discrepancy factor (adf) for O$^{2+}$ reaching 89 -- the second highest known value after that found for the hydrogen deficient knots in Abell~30.  The observed temperature diagnostics and abundances support the idea that, like Abell~30, the knot of Abell~58 contains some very cold ionised material.  Although the central star is carbon-rich (C/O$>$1), the knot is found to be oxygen-rich, a situation not predicted by the single-star `born again' theory of its formation.

We compare the known properties of Abell~58 to those of Abell~30, Sakurai's Object and several novae and nova remnants.  We argue that the abundances in the ejecta observed in A\,30 and A\,58 have more in common with neon novae than with Sakurai's Object, which is believed to have undergone a final helium flash.  In particular, the C/O ratio of less than unity and presence of substantial quantities of neon in the ejecta of both Abell~30 and Abell~58 are not predicted by very late thermal pulse models.

\end{abstract}
 

\section{Introduction}

\begin{figure}
 \epsfig{file=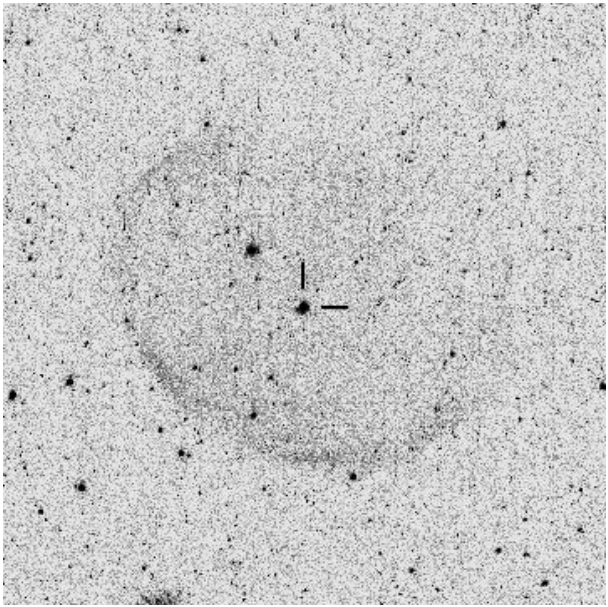, width=8cm, height=8cm, angle=0}
 \caption{Hubble Space Telescope image of Abell 58 (WFPC2, F658N filter isolating [N~{\sc ii}] emission), showing the faint outer nebulosity and bright central knot.  North is up and east is to the left, and the field of view is approximately 70 arcsec on each side.  The position marks indicate the central knot; the point source within the nebula to the north-east is a field star.}
 \label{A58fig}
\end{figure}

Abell~58 (V605~Aql) consists of a large (44 $\times$ 36 arcsec) faint shell, with a brighter knot at its geometric centre (Figure~\ref{A58fig}).  The knot is assumed to have formed in a nova-like outburst which peaked in 1919, by which time the central star had increased in brightness by five magnitudes.  A\,58 is one of a very small number of objects which have been seen to undergo rapid stellar evolution, and is considered an `older twin' of Sakurai's object (V4334 Sgr).  Since the outburst, the central star has become heavily obscured, and its spectrum is now only visible in scattered light (Clayton et al. 2006).

The central knot is extremely hydrogen-deficient (Seitter 1985, Guerrero \& Manchado 1996), and high-resolution spectroscopy by Pollacco et al. (1992) showed that its emission was blue-shifted relative to the main nebula by $\sim$100\,km\,s$^{-1}$.  They proposed that the knot was one side of a bipolar collimated flow, the other side being obscured by dust.  The spectrum of A\,58's central knot has been studied by Guerrero \& Manchado (1996, hereafter GM96), who found He/H to be 1.24 and O/H to be 1.25$\times$10$^{-2}$, by number.  However, they considered this value of O/H to be unreliably high due to probable shock excitation.

V605~Aql itself cannot be seen directly, as it is hidden by a thick dust torus, but stellar emission lines are seen via scattered light.  Clayton et al. (2006) derived surface abundances for the star from observations of its scattered light, finding them to be typical of Wolf-Rayet central stars of PNe, with $\sim$55\% helium and $\sim$40\% carbon by mass.  They estimated the star's surface temperature to be 95\,000\,K.

Abell 30 is a nebula containing similar but much older hydrogen-deficient knots, ejected from its central star $\sim$1000 years ago (Reay, Atherton \& Taylor 1983).  The knots of Abell\,30 have been studied in detail by Wesson, Liu \& Barlow (2003) and Ercolano et al (2003).  Empirical analysis and subsequent photoionisation modelling of the polar knots showed that they contain extremely cold ($\lesssim$2000\,K) ionised material.  The existence of knots of cold ionised material in `normal' planetary nebulae has been proposed as a way of reconciling the ubiquitous discrepancy whereby abundances derived from optical recombination lines (ORLs) are higher than those derived from collisionally excited lines (CELs) (Liu et al. 2000).  The `normal' nebula which shows the most extreme abundance discrepancy factor (that is, the ratio of the abundance of an element derived from recombination lines to that derived from collisionally excited lines) is Hf\,2--2, for which the oxygen abundance derived from ORLs is 68 times higher than the value derived from ORLs (Liu et al. 2006).   This is intermediate between the typical discrepancy for oxygen of 3-5 (Wesson et al 2005) and the extreme value of $\sim$700 seen in A\,30 (Wesson et al. 2003).

The generally accepted explanation of the origin of the knots in A\,30 and A\,58 is that the central star, after the formation of its surrounding planetary nebula, underwent a very late thermal pulse (VLTP), which ejected freshly processed stellar material into the centre of the nebula (Iben et al. 1983, Herwig 2001).  However, the analysis of Abell~30's polar knots showed that they were oxygen-rich (Wesson et al. 2003), the opposite of the situation predicted by the born-again theory.  The observed C/O ratios are similar to those seen in PNe with large abundance discrepancies, such as NGC\,6153 (Liu et al. 2000).

This paper presents an analysis of long-slit spectra of A\,58's central knot.  We find that its properties are very similar to those found for A\,30's knots and consistent with the Ercolano et al. (2003) model whereby a cold, extremely hydrogen-deficient core contributes essentially all the strong ORL emission, and a hot outer shell emits CELs.  As with A\,30, the knots are found to be oxygen-rich.  We discuss the implications of these findings for the born-again theory, and for planetary nebulae in general.

\section{Observations}

A\,58 was observed using the double-armed ISIS spectrograph on the 4.2-m William Herschel Telescope (WHT) at the Observatorio del Roque de los Muchachos, on La Palma, Spain, on the night of 2003 August 01.  The seeing was sub-arcsecond throughout the observations.  The spectrograph slit was placed over the central knot of A\,58 at a position angle of 237$^\circ$, matching that of the major axis of the outer nebula.  The slit width of 0.78 arcsec should have caught all the flux from the knot, which is about 0.7 arcsec wide (Clayton \& De Marco 1997).  Spectra covering wavelengths from 3500 to 5100 {\AA} and 5100 to 7600 {\AA} were taken, with R600B and R300R gratings giving spectral resolutions of 1.45{\AA} and 2.65{\AA} on the blue and red arms respectively.  Nine half-hour exposures were taken, for a total exposure time in each arm of 4.5 hours.

The data were reduced using standard procedures in the {\sc midas} package {\sc long92}\footnote{{\sc midas} is developed and distributed by the European Southern Observatory.} and {\sc IRAF}\footnote{{\sc iraf} is distributed by the National Optical Astronomy Observatories.}.  They were bias-subtracted, flat-fielded, and wavelength-calibrated using a Cu-Ne lamp for the red spectra and a Cu-Ar lamp for the blue spectra.  The observations were flux-calibrated using observations of the standard star Hz\,44.

\begin{figure}
 \epsfig{file=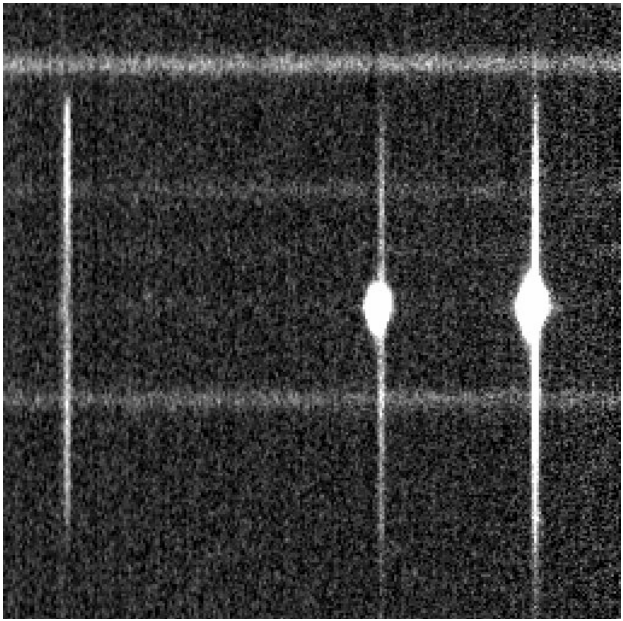, width=8.2cm}
 \caption{Part of a two-dimensional frame showing emission from the knot and outer nebula of Abell~58.  The three lines visible are H$\beta$, [O~{\sc iii}] $\lambda$4959 and [O~{\sc iii}] $\lambda$5007.  The knot emits strongly in [O~{\sc iii}] but H$\beta$ emission is small compared to that from the outer nebula.}
 \label{2dframe}
\end{figure}

\begin{figure*} 
 \epsfig{file=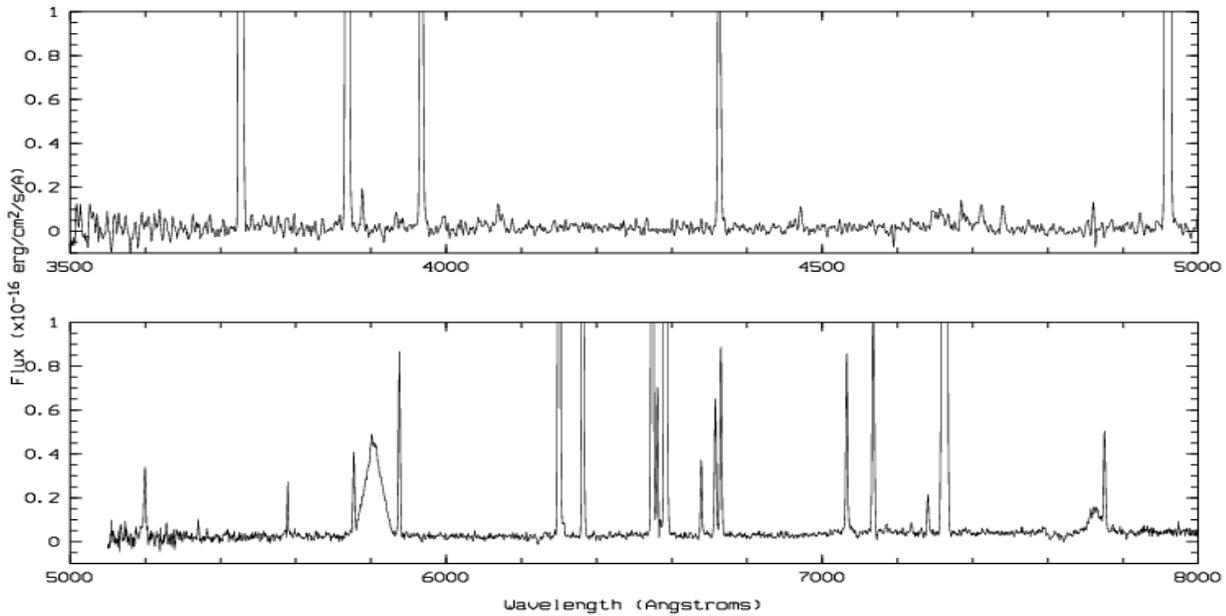, width=17cm, height=9cm, angle=0}
 \caption{Spectrum of the central knot of Abell~58.  Note the prominent nebular recombination line features at $\sim$4075 and $\sim$4650{\AA}, and the strong stellar C~{\sc iv} emission at 5801,5812{\AA} and 7377{\AA}.}
 \label{spectrum}
\end{figure*}

The knot shows strong emission from both CELs and heavy element recombination lines.  A number of emission lines from the fainter outer nebula are clearly visible across the spectrograph slit (Figure~\ref{2dframe}).  When extracting the spectrum of the central knot, the contribution of the outer nebula was subtracted using a spectrum extracted from a region outside the knot covering the same number of rows on the CCD chip as the knot spectrum.  A spectrum of the outer nebula was also extracted, excluding the rows on the CCD chip covered by the central knot.  The spectrum of the central knot is shown in Figure~\ref{spectrum}.

Lines in the spectra of the knot and outer nebula were identified and measured by fitting Gaussian profiles in {\sc midas}.   Line fluxes are conventionally normalised to H$\beta$=100 in nebular studies, but in this case as H$\beta$ is so weak, to avoid large numbers line fluxes are normalised such that I(H$\beta$)=1.  A list of lines measured in the knot and outer nebula is presented in Table~\ref{A58linelist}.  We measure a total H$\beta$ flux from the knot of 3.8$\times$10$^{-17}$\,erg\,cm$^{-2}$\,s$^{-1}$.  Lines in the central knot have an FWHM equivalent to $\sim$190\,km\,s$^{-1}$ after correction for instrumental broadening, while the lines from the outer nebula have an FWHM equivalent to $\sim$85\,km\,s${-1}$.  These values are consistent with those derived from high-resolution echelle spectra by Pollacco et al. (1992).

The broadening of lines from the knot makes the detection and accurate measurement of weak recombination lines much more difficult than is the case for the narrow lines of Abell~30.  The difficulty is most acute in regions where many lines are blended, such as from 4068--4075 {\AA}, and the $\lambda$4650 complex of O~{\sc ii} lines.  Multiple gaussians fitted to the broad features at these wavelengths give line fluxes for the brightest lines, but with fairly significant errors.  Overall we detect 42 lines emitted by the knot, due to H, He, C, N, O, Ne, Ar, S, Cl and Fe, rather less than the $>$100 lines seen in our spectra of knots J1 and J3 in Abell~30 (Wesson et al. 2003).  None the less, these spectra are the deepest that have been obtained of this object, and are of high enough quality to carry out a thorough analysis of the properties of the knot.  We find typical errors on our flux measurements of 40-55\% for lines with F($\lambda$)$<$0.2, where F(H$\beta$)\,=\,1, 10-40\% for lines with 0.2$\le$F($\lambda$)$\le$3.0, and $<$5\% for lines with F($\lambda$)$>$3.0.

\begin{table*}
\centering
\label{A58linelist}
\caption{Line fluxes measured from the spectrum of the knot and outer nebula of Abell~58.  Fluxes are normalised such that F(H$\beta$)=1.  We used $c({\rm H}\beta)$=2.0 for the knot and $c({\rm H}\beta)$=1.04 for the outer nebula (Section~\ref{diagnostics}) to derive the dereddened intensities, I($\lambda$).}
\begin{tabular}{ccccccccccccc}

\hline

\multicolumn{3}{c}{Knot}&\multicolumn{3}{c}{Background nebula}& & & & & & & \\
$\lambda_{\rm obs}$&$F(\lambda)$&$I(\lambda)$&$\lambda_{\rm obs}$&$F(\lambda)$&$I(\lambda)$&Ion&$\lambda_{\rm 0}$&Mult&Lower Term&Upper Term&$g_1$&$g_2$\\

\hline

3725.34 & 62.31 & 201.9 & 3726.69 & 0.453 & 0.842 & [O II]   & 3726.03 & F1   &   2p3 4S*  &   2p3 2D* & 4 &  4\\
3728.12 & 48.37 & 156.6 & 3729.47 & 0.807 & 1.497 & [O II]   & 3728.82 & F1   &   2p3 4S*  &   2p3 2D* & 4 &  6\\
3868.21 & 126.2 & 356.1 & 3869.53 & 0.235 & 0.406 & [Ne III] & 3868.75 & F1   &   2p4 3P   &   2p4 1D  & 5 &  5\\
3888.55 & 1.624 & 4.496 & *       & *     & *     & He I     & 3888.65 & V2   &   2s  3S   &   3p  3P* & 3 &  9\\
3967.01 & 43.51 & 111.7 & *       & *     & *     & [Ne III] & 3967.46 & F1   &   2p4 3P   &   2p4 1D  & 3 &  5\\
4068.97 & 0.852 & 1.986 & *       & *     & *     & O II     & 4069.62 & V10  &   3p  4D*  &   3d  4F  & 2 &  4\\
 *      & *     & *     & *       & *     & *     & O II     & 4069.89 & V10  &   3p  4D*  &   3d  4F  & 4 &  6\\
 *      & 0.631 & 1.472 & *       & *     & *     & [S II]   & 4068.60 & F1   &   2p3 4S*  &   2p3 2P* & 4 & 4\\
4075.01 & 0.614 & 1.423 & *       & *     & *     & O II     & 4075.86 & V10  &   3p  4D*  &   3d  4F  & 8 & 10\\
 *      & 0.227 & 0.525 & *       & *     & *     & [S II]   & 4076.35 & F1   &   2p3 4S*  &   2p3 2P* & 2 & 4\\
4087.82 & 0.410 & 0.939 & *       & *     & *     & O II     & 4089.29 & V48a &   3d  4F   &   4f  G5* & 10& 12\\
 *      & *     & *     & 4103.11 & 0.125 & 0.192 & H 6      & 4101.74 & H6   &   2p+ 2P*  &   6d+ 2D  & 8 & 72\\
4266.84 & 0.521 & 1.002 & *       & *     & *     & C II     & 4267.15 & V6   &   3d  2D   &   4f  2F* & 10& 14\\
4339.21 & 0.448 & 0.799 & 4341.81 & 0.331 & 0.449 & H 5      & 4340.47 & H5   &   2p+ 2P*  &   5d+ 2D  & 8 & 50\\
4363.01 & 20.41 & 35.52 & *       & *     & *     & [O III]  & 4363.21 & F2   &   2p2 1D   &   2p2 1S  & 5 &  1\\
4471.38 & 1.211 & 1.878 & *       & *     & *     & He I     & 4471.50 & V14  &   2p  3P*  &   4d  3D  & 9 & 15\\
4647.28 & 0.691 & 0.879 & *       & *     & *     & O II     & 4649.13 & V1   &   3s  4P   &   3p  4D* & 6 & 8 \\
4648.99 & 0.404 & 0.514 & *       & *     & *     & O II     & 4650.84 & V1   &   3s  4P   &   3p  4D* & 2 & 2 \\
4655.70 & 0.780 & 0.992 & *       & *     & *     & [Fe III] & 4658.10 & F3   &   3d6 5D   &   3d6 3F2 & 9 &  9\\
4660.04 & 0.692 & 0.869 & *       & *     & *     & O II     & 4661.63 & V1   &   3s  4P   &   3p  4D* & 4 &  4\\
4667.68 & 0.720 & 0.897 & *       & *     & *     & O II     & 4669.27 & V89b &   3d  2D   &   4f  D2* & 4 &  6\\
4685.06 & 1.310 & 1.597 & *       & *     & *     & He II    & 4685.68 & 3.4  &   3d+ 2D   &   4f+ 2F* & 18 & 32\\
4711.61 & 1.954 & 2.312 & *       & *     & *     & [Ar IV]  & 4711.37 & F1   &   3p3 4S*  &   3p3 2D* & 4 &  6\\
4723.56 & 0.221 & 0.258 & *       & *     & *     & [Ne IV]  & 4724.15 & F1   &   2p3 2D*  &   2p3 2P* & 4 &  4\\
 *      & *     & *     & *       & *     & *     & [Ne IV]  & 4725.62 & F1   &   2p3 2D*  &   2p3 2P* & 4 &  2\\
4740.46 & 1.731 & 1.981 & *       & *     & *     & [Ar IV]  & 4740.17 & F1   &   3p3 4S*  &   3p3 2D* & 4 &  4\\
4860.60 & 1.000 & 1.000 & 4863.00 & 1.000 & 1.000 & H 4      & 4861.33 & H4   &   2p+ 2P*  &   4d+ 2D  & 8 & 32\\
4922.86 & 0.929 & 0.862 & *       & *     & *     & He I     & 4921.93 & V48  &   2p  1P*  &   4d  1D  & 3 &  5\\
4958.96 & 366.5 & 326.6 & 4960.64 & 0.717 & 0.674 & [O III]  & 4958.91 & F1   &   2p2 3P   &   2p2 1D  & 3 &  5\\
5006.79 & 1123  & 948.4 & 5008.44 & 1.978 & 1.808 & [O III]  & 5006.84 & F1   &   2p2 3P   &   2p2 1D  & 5 &  5\\
5198.91 & 11.48 & 7.899 & *       & *     & *     & [N I]    & 5199.84 & F1   &   2p3 4S*  &   2p3 2D* & 4 &  4\\
5340.93 & 1.486 & 0.890 & *       & *     & *     & C II     & 5342.38 &      &   4f  2F*  &   7g  2G  & 14 & 18\\
5754.48 & 10.38 & 4.290 & *       & *     & *     & [N II]   & 5754.60 & F3   &   2p2 1D   &   2p2 1S  & 5 &  1\\
5875.62 & 20.48 & 7.682 & 5877.53 & 0.305 & 0.182 & He I     & 5875.66 & V11  &   2p  3P*  &   3d  3D  & 9 & 15\\
6300.14 & 225.4 & 62.21 & 6301.45 & 0.268 & 0.136 & [O I]    & 6300.34 & F1   &   2p4 3P   &   2p4 1D  & 5 &  5\\
6363.63 & 78.64 & 20.81 & 6364.22 & 0.283 & 0.140 & [O I]    & 6363.78 & F1   &   2p4 3P   &   2p4 1D  & 3 &  5\\
6547.83 & 172.8 & 40.63 & 6549.92 & 2.420 & 1.127 & [N II]   & 6548.10 & F1   &   2p2 3P   &   2p2 1D  & 3 &  5\\
6561.79 & 17.55 & 4.090 & 6564.75 & 6.119 & 2.834 & H 3      & 6562.77 & H3   &   2p+ 2P*  &   3d+ 2D  & 8 & 18\\
6583.21 & 523.0 & 120.2 & 6585.36 & 6.870 & 3.160 & [N II]   & 6583.50 & F1   &   2p2 3P   &   2p2 1D  & 5 &  5\\
6678.27 & 9.573 & 2.076 & *       & *     & *     & He I     & 6678.16 & V46  &   2p  1P*  &   3d  1D  & 3 &  5\\
6715.99 & 16.49 & 3.494 & 6718.47 & 1.182 & 0.521 & [S II]   & 6716.44 & F2   &   2p3 4S*  &   2p3 2D* & 4 &  6\\
6730.52 & 23.35 & 4.903 & 6732.80 & 0.849 & 0.372 & [S II]   & 6730.82 & F2   &   2p3 4S*  &   2p3 2D* & 4 &  4\\
7065.27 & 22.48 & 3.881 & *       & *     & *     & He I     & 7065.25 & V10  &   2p  3P*  &   3s  3S  & 9 &  3\\
7135.75 & 42.51 & 7.059 & 7137.93 & 0.310 & 0.120 & [Ar III] & 7135.80 & F1   &   3p4 3P   &   3p4 1D  & 5 &  5\\
7236.42 & *     & 0.199 & *       & *     & *     & C II     & 7236.42 & V3   &   3p  2P*  &   3d  2D  & 4 &  6\\
        & *     & *     & *       & *     & *     & C II     & 7237.17 & V3   &   3p  2P*  &   3d  2D  & 4 &  4\\
7281.16 & 4.321 & 0.662 & *       & *     & *     & He I     & 7281.35 & V45  &   2p  1P*  &   3s  1S  & 3 &  1\\
7319.81 & 174.3 & 26.20 & *       & *     & *     & [O II]   & 7318.92 & F2   &   2p3 2D*  &   2p3 2P* & 6 &  2\\
 *      & *     & *     & *       & *     & *     & [O II]   & 7319.99 & F2   &   2p3 2D*  &   2p3 2P* & 6 &  4\\
7330.07 & 148.2 & 22.15 & *       & *     & *     & [O II]   & 7329.67 & F2   &   2p3 2D*  &   2p3 2P* & 4 &  2\\
 *      & *     & *     & *       & *     & *     & [O II]   & 7330.73 & F2   &   2p3 2D*  &   2p3 2P* & 4 &  4\\
8045.82 & 1.812 & 0.191 & *       & *     & *     & [Cl IV]  & 8045.63 &      &   3p2 3P   &   3p2 1D  & 5 &  5\\
\hline
\end{tabular}
\end{table*}

\section{Outer nebula}

We detect only the brightest lines from the outer nebula of Abell~58.  We derive a logarithmic extinction at H$\beta$, $c({\rm H}\beta)$ of 1.04 from the observed H$\alpha$/H$\beta$ ratio.  This is similar to the value obtained using the reddening maps of Schlegel et al. (1998), which for A\,58's galactic longitude and latitude of 37.60$^\circ$ and -5.16$^\circ$ respectively give E(B-V)=0.544.  This is equivalent to A$_v$=1.69, and $c({\rm H}\beta)$=1.16.

No temperature diagnostics are available, and both the [O~{\sc ii}] and [S~{\sc ii}] density diagnostic line ratios fall below the low density limit.  We calculate ionic abundances for three temperatures -- 7.5, 10 and 15\,kK -- assuming that the density is 200\,cm$^{-3}$, the value found by GM96.  Total abundances are calculated using the ionisation correction scheme of Kingsburgh \& Barlow (1994).  The results are shown in Table~\ref{outerabunds}.  The spectrum of the outer nebula is of low excitation, with I([O~{\sc ii}] 3727+3729)~$\sim$I([O~{\sc iii}] 4959+5007), so a T$_{\rm e}$ between 7.5 and 10kK seems most likely.

We derive a Ne/O ratio between 0.49 and 0.74, depending on the temperature adopted.  This is rather higher than the average value for PNe of about 0.25.  At 7.5kK, the derived oxygen abundance (on a logarithmic scale where N(H)=12.0) of 8.74 is close to the average value of 8.68 found for galactic PNe by Kingsburgh \& Barlow (1994), while the neon abundance of 8.61 is more than a factor of three higher than their average.  At 10\,kK, on the other hand, the derived neon abundance is about 30 per cent lower than the average found by Kingsburgh \& Barlow (1994), while the oxygen abundance is a factor of three lower.

Assuming T$_{\rm e}$=7.5\,kK gives N/O~=~0.38, while T$_{\rm e}$=10\,kK gives N/O~=~0.68.  Adopting a temperature of 15\,kK gives N/O~=~1.20, which would classify A\,58 as a Type~I planetary nebula as defined by Kingsburgh \& Barlow (1994), with N/O$>$0.8.  The criteria of Peimbert \& Torres-Peimbert (1983) require N/O$>$0.5 and He/H$>$0.125, and so the outer nebula of A\,58 would be classified as a Peimbert Type~I nebula if its temperature is higher than about 8.5\,kK.

\begin{table}
\centering
\caption{Elemental abundances in the outer shell of Abell~58}
\label{outerabunds}
\begin{tabular}{lllll}
\hline
Abundance       & \multicolumn{3}{c}{Temperature (kK)} \\
                & 7.5   & 10    & 15    \\
\hline
He$^+$/H$^+$                 & 0.125 & 0.133 & 0.141 \\
He$^{2+}$/H$^+$              & -     & -     &       \\
\vspace{0.1cm}
He/H                         & 0.125 & 0.133 & 0.141 \\
N$^+$/H$^+\times$10$^{-5}$   & 12.7  & 5.51  & 2.26 \\
ICF                          & 1.62  & 1.84  & 2.11 \\
\vspace{0.1cm}
N/H $\times$10$^{-5}$        & 20.6  & 10.1  & 4.77 \\
O$^+$/H$^+\times$10$^{-5}$   & 33.7  & 8.18  & 1.87 \\
O$^{2+}$/H$^+\times$10$^{-5}$& 21.0  & 6.88  & 2.07 \\
ICF                          & 1.00  & 1.00  & 1.00 \\
\vspace{0.1cm}
O/H$\times$10$^{-5}$         & 54.7  & 15.06 & 3.94 \\
Ne$^{2+}$/H$^+\times$10$^{-5}$&15.6  & 4.09  & 1.02 \\
ICF                          & 2.60  & 2.19  & 1.90 \\
Ne/H$\times$10$^{-5}$        & 40.56 & 8.96  & 1.94 \\
\hline
\end{tabular}
\end{table}

\section{Central knot}

\subsection{General properties}

The knot of Abell\,58 is expanding at about 100\,km\,s$^{-1}$, and is moving at -100\,km\,s$^{-1}$ relative to the systemic velocity (Pollacco et al. 1992).  Estimates of the distance to the system have ranged from $\sim$3\,kpc (Cahn \& Kaler 1971) to 6\,kpc (Pollacco et al. 1992).  The knot is ionised by a star which is completely obscured from view, although stellar emission lines are visible in scattered light.  Clayton et al. (2006) determined a stellar temperature of 95\,kK and a luminosity of 10$^4$L$_{\sun}$, assuming a distance of 3.5\,kpc.

We estimate the mass of the knot in two ways.  First, we use the following relation:

\begin{equation}
M_k~=\frac{4\pi D^2I(He I \lambda5876)\mu_{He}}{n_e\alpha_{eff(\lambda5876)}E_{\lambda5876}},
\end{equation}

where D is the distance to the knot, I(He {\sc i} $\lambda$5876) is the dereddened flux of the He~{\sc i} line at 5876{\AA}, $\mu_{He}$ is the mean ionic mass per He$^+$ ion, $n_{\rm e}$ is the electron density, $\alpha_{eff(5876)}$ is the effective recombination coefficient of the $\lambda$5876 line and E$_{\lambda5876}$ is the energy of each photon at 5876{\AA}.  Adopting a distance of 3.5\,kpc from Clayton et al. (2006), an electron density of 2100\,cm$^{-3}$ (Section~\ref{diagnostics}) and a mean ionic mass per He$^+$ ion of 9.2, based on ionic abundances derived from ORLs (Section~\ref{abundances}), we derive a mass of 5.25$\times$10$^{-5}$M$_{\sun}$ for the knot.

Second, we derive a mass based on the angular size of the knot:

\begin{equation}
M_k~=~\frac{4\pi}{3}\Theta^3D^3\mu_e\epsilon n_e,
\end{equation}

where $\Theta$ is the angular radius of the knot, D its distance, $\mu_e$ the mean ionic mass per electron, $\epsilon$ the volume filling factor, and $n_{\rm e}$ the electron density.  We measure the angular radius of the knot from Hubble Space Telescope images which we downloaded from the STSci archive.  The images were taken in 2001 as part of GO program 9092, in narrow-band filters isolating [O~{\sc iii}] and [N~{\sc ii}] emission.  We derive an angular radius of 0.38 arcsec.  Taking D to be 3.5\,kpc, and $\mu_e$ to be 6.16, based on abundances derived from ORLs (Section~\ref{abundances}), we derive a mass of 3.22$\times$10$^{-4}$$\epsilon\,$M$_{\sun}$.  Equating these two masses we find that $\epsilon$~=~0.16, if $n_{\rm e}$~=~2100\,cm$^{-3}$.

\subsection{Extinction}

GM96 found a value of $c({\rm H}\beta)$ of 0.29 for the outer nebula of A\,58, but with a large uncertainty due to the weakness of the hydrogen line intensities.  They suggested that the extinction in the central knot could be higher.  Most of the helium in the knot of A\,58 is singly ionised (see Section~\ref{Heabunds}) but the weakness of the hydrogen lines means that the contribution of the He~{\sc ii} Pickering series lines to H$\alpha$ and H$\beta$ lines needs to be corrected for when calculating the extinction.

The He~{\sc ii} line at $\lambda$4859 is not seen in our blue spectrum, although the spectral resolution is high enough to resolve it from H$\beta$.  We calculate from its theoretical ratio to He~{\sc ii} $\lambda$4686 that its strength should be $\sim$6\% of that of the feature at H$\beta$.  We find the actual value of $c({\rm H}\beta)$ by iteratively calculating $c({\rm H}\beta)$ and then the strength of the $\lambda$6561 line for that value of $c({\rm H}\beta)$.  We find that the H$\alpha$/H$\beta$ ratio is 16.10, giving $c({\rm H}\beta)$ of 2.38.  However, the extreme weakness of the observed H$\alpha$ and H$\beta$ lines means that this value is very uncertain.

\begin{figure}
 \epsfig{file=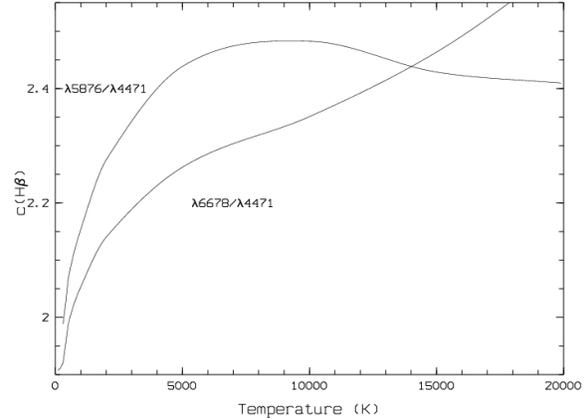, width=8cm, angle=0}
 \caption{The range of values of $T_{\rm e}$ and $c({\rm H}\beta)$ which can lead to the observed values of He~{\sc i} recombination line ratios.}
\label{He_c} 
\end{figure}

We also measured the extinction using the relative intensities of He~{\sc i} recombination lines.  The observed line ratios $\lambda$5876/$\lambda$4471 and $\lambda$6678/$\lambda$4471 depend both on the temperature of the gas and the extinction, and so in principle the two line ratios can be used to derive the temperature and extinction simultaneously.  In Figure~\ref{He_c} we plot the possible values of $T_{\rm e}$ and $c({\rm H}\beta)$ which could give rise to our observed line ratios.  The two lines intersect at $c({\rm H}\beta)$=2.44 and $T_{\rm e}$=14\,000\,K, and also lie very close to each other at $c({\rm H}\beta)$=2.0 and $T_{\rm e}$$<$1000\,K.  Given the very high abundance of CNO coolants we derive from ORLs (Section~\ref{ORLabunds}), a temperature of 14\,000\,K for the region in which the helium lines are emitted seems implausibly high.  Three-dimensional photoionisation modelling of the knots of Abell~30 by Ercolano et al. (2003) indicated that in that case helium emission came predominantly from the very cold core of the knot.  We therefore believe that the cold solution is physically more realistic for the knot of Abell~58, and adopt $c({\rm H}\beta)$=2.0 for our subsequent analysis.

We note that this value, as well as the value of 1.04 derived for the outer nebula, is much higher than that adopted by GM96 for their analysis.  Their value was derived from the outer nebula alone, no H$\beta$ flux being measured from the knot.  They noted that the weakness of the lines made this value subject to large errors, and that there was evidence of higher extinction in the central regions.  Our deeper spectra allow a more accurate determination of $c({\rm H}\beta)$ in the outer nebula, giving a value consistent with the Schlegel et al. (1998) reddening maps, and confirm the expected higher extinction in the central knot.

\subsection{Temperature and density}
\label{diagnostics}

The electron density of the central knot was measured from the [O~{\sc ii}] $\lambda\lambda$3726,2729, [S~{\sc ii}] $\lambda\lambda$6717,6731 and [Ar~{\sc iv}] $\lambda\lambda$4711,4740 line ratios, which give 1520, 2730 and 2050\,cm$^{-3}$ respectively.  We adopt a density of 2100\,cm$^{-3}$.  This value is expected to be representative of the hotter regions of the knot, from which essentially all of the CEL emission will arise.  Recent atomic data for O$^{2+}$ (Bastin et al. 2006) allows electron density determinations based on the relative intensities of the lines in the O~{\sc ii} V1 multiplet, which could be emitted predominantly by a cold core, but unfortunately we do not see enough V1 lines to use this technique to derive a density for the cold parts of the knot.

The weak auroral lines of [O~{\sc iii}] and [N~{\sc ii}] are not visible in our spectrum of the outer nebula, but both are well detected in the knot.  The temperatures derived from the [O~{\sc iii}] and [N~{\sc ii}] nebular-to-auroral line ratios in the knot are 20,800\,K and 15,200\,K respectively.  The [O~{\sc ii}] $\lambda$7320,7330/$\lambda$3726,3729 line ratio gives an unphysically high temperature.  The large difference in wavelength between these two sets of lines makes this diagnostic very sensitive to errors in the adopted value of $c({\rm H}\beta)$, and the critical densities also differ widely.

CEL temperature diagnostic ratios may be significantly affected by recombination excitation (Rubin 1986).  The amount of O$^{3+}$ in the nebula is expected to be negligible, and therefore recombination excitation will not affect the temperature derived from the [O~{\sc iii}] lines.  However, most nitrogen will be in the form of N$^{2+}$, and therefore we use equation 1 from Liu et al. (2000) to calculate the contribution of recombination to the observed [N~{\sc ii}] $\lambda$5754 line.  Estimating N$^{2+}$/H$^+$ from the observed CEL N$^+$/H$^+$ abundance (Section~\ref{abundances}) and ICFs from Kingsburgh and Barlow (1994), we find that the recombination contribution is negligible.  A recombination line N$^{2+}$/H$^+$ abundance is not available, but even if the ORL abundance of N$^{2}$/H$^+$ was two orders of magnitude greater than the CEL abundance, the correction to the temperature would only amount to a reduction of 700\,K.  Therefore we assume that the $\lambda$5754 line is produced purely by collisional excitation, and in our subsequent abundance analysis we use a temperature of 15\,200\,K to derive CEL abundances from singly ionised species, and 20\,800\,K for more highly ionised species.  These very high electron temperatures suggest that grain photoelectron heating is the dominant heating mechanism in the hot regions of the knot, by analogy with IRAS 18333-2357 (Borkowski \& Harrington 1991) and Abell\,30 (Ercolano et al. 2003). 

We also determined temperatures from the ratios of helium recombination lines.  These vary weakly with temperature (Smits 1996), and in previous large sample studies have been found to be systematically lower than temperatures measured from the hydrogen Balmer jump and CEL diagnostics, but higher than those derived from O~{\sc ii} recombination lines (e.g. Wesson et al 2005).  We find that the ratio of He~{\sc i} $\lambda$5876 to $\lambda$4471 implies a temperature of 350\,K, while the ratio of He~{\sc i} $\lambda$6678 to $\lambda$4471 implies a temperature of 550\,K.  We adopt T(He~{\sc i})~=~500\,K.  These values are much lower than the values of 8\,850\,K and 4\,450\,K found for the knots of A\,30.

Finally, we estimate the temperature in the knot based on the weak temperature dependence of O~{\sc ii} recombination line ratios.  The ratios of some O~{\sc ii} lines (e.g. $\lambda$4075/$\lambda$4089, $\lambda$4649/$\lambda$4089) are weakly sensitive to temperature, and have been used to derive temperatures in several planetary nebulae (e.g. Wesson et al. 2003, Tsamis et al. 2004).  However, these derivations were based on atomic data from Storey et al. 1994 and Liu et al. 1995, which assumed a statistical distribution of population in the three ground levels of O$^{2+}$, $^3$P$_0^{\rm o}$, $^3$P$_1^{\rm o}$ and $^3$P$_2^{\rm o}$.  The populations of these levels are sensitive to electron density at typical nebular densities, and Bastin \& Storey (2006) and Storey (2007) have recently calculated recombination coefficients accounting for this effect.  Their results show that the line ratios vary significantly with density: for example, an observed value of 2.5 for the $\lambda$4649/$\lambda$4089 ratio may indicate a temperature of $\sim$1000\,K at a density of 10$^4$cm$^{-3}$, or $\sim$5000\,K at 10$^2$cm$^{-3}$ (Storey 2007).  In A\,58, we find a ratio of 0.94, which is lower than the value derived for any density at 600\,K by Storey (2007), providing strong evidence that the knot of A\,58 contains ionised material at very low temperatures.  This indicates that the situation in the knot of Abell~58 is similar to that seen in Abell~30, with cold, metal-rich ionised plasma existing within the knot.  We adopt an electron temperature of 500\,K for our derivation of abundances from ORLs.

Our $T_{\rm e}$ and $n_{\rm e}$ diagnostics are summarised in Table~\ref{diagtable}.

\begin{table}
\centering
\caption{Temperature and density diagnostics for the central knot of Abell 58}
\label{diagtable}
\begin{tabular}{lll}
\hline
Diagnostic ratio & Value & Temperature (K) \\
\hline
~[O~{\sc iii}]($\lambda$4959+$\lambda$5007)/$\lambda$4363 & 35.9 & 20\,800 \\
~[O~{\sc ii}] ($\lambda$3727+$\lambda$3729)/($\lambda$7320+$\lambda$7330) & 4.18 & $>$20\,000 \\
~[N~{\sc ii}] ($\lambda$6548+$\lambda$6584)/$\lambda$5754 & 37.5 & 15\,200 \\
~He~{\sc i} $\lambda$5876/$\lambda$4471 & 4.09 & 350 \\
~He~{\sc i} $\lambda$6678/$\lambda$4471 & 1.11 & 550 \\
~O~{\sc ii} $\lambda$4649/$\lambda$4089 & 0.94 & $<$2\,000 \\
~O~{\sc ii} $\lambda$4075/$\lambda$4089 & 1.52 & $<$2\,000 \\
\hline
Diagnostic ratio & Value & Density (cm$^{-3}$) \\
\hline
~[O~{\sc ii}] $\lambda$3729/$\lambda$3726 & 0.775 & 1520 \\
~[S~{\sc ii}] $\lambda$6717/$\lambda$6731 & 0.713 & 2730 \\
~[Ar~{\sc iv}] $\lambda$4711/$\lambda$4740 & 1.167 & 2050 \\
\hline
\end{tabular}
\end{table}

\subsection{Ionic abundances from CELs}
\label{CELabunds}

To derive chemical abundances in Abell~58's central knot, we follow the same general approach as described in detail by Wesson et al. (2003), using the same atomic data.  Ionic abundances were derived from collisionally excited lines using a temperature of 15\,200\,K for singly-ionised species, and 20\,800\,K for more highly ionised species, except for the several ions for which atomic data are only available for temperatures up to 20\,000\,K.  In these cases, ([Ar~{\sc iii}], [Ar~{\sc iv}] and [S~{\sc iii}]), a temperature of 20\,000\,K was adopted, and the abundance thus derived would be expected to be higher than the true value.  The abundances derived from CELs are presented in Table~\ref{CELtable}.

\begin{table}
\centering
\caption{Ionic abundances derived from CELs}
\label{CELtable}
\begin{tabular}{lll}
\hline
Species & Lines & Abundance \\
\hline
N$^+$/H$^+$     & $\lambda\lambda$6548,6584 & 8.26$\times$10$^{-4}$\\ 
O$^+$/H$^+$     & $\lambda\lambda$3726,3729 & 3.70$\times$10$^{-3}$\\ 
O$^+$/H$^+$     & $\lambda\lambda$7319,7330 & 7.17$\times$10$^{-3}$\\ 
O$^{2+}$/H$^+$  & $\lambda\lambda$4959,5007 & 5.16$\times$10$^{-3}$\\ 
Ne$^{2+}$/H$^+$ & $\lambda\lambda$3868,3967 & 4.01$\times$10$^{-3}$\\ %
Ne$^{3+}$/H$^+$ & $\lambda$4725             & 1.94$\times$10$^{-4}$\\ %
S$^+$/H$^+$     & $\lambda\lambda$6716,6731 & 1.03$\times$10$^{-5}$\\ 
Ar$^{2+}$/H$^+$ & $\lambda$7135             & 1.73$\times$10$^{-5}$\\ 
Ar$^{3+}$/H$^+$ & $\lambda\lambda$4711,4740 & 8.51$\times$10$^{-6}$\\ %
\hline
\end{tabular}
\end{table}

\subsection{Ionic abundances from ORLs}
\label{ORLabunds}

Given the exponential temperature dependence of CEL emissivities, and the inverse power-law temperature dependence of ORL emissivities, strongly enhanced ORL emission provides evidence in favour of a cold, extremely hydrogen-deficient core in A\,58's knot.  Observations and modelling supported this interpretation in the case of the polar knots of Abell~30 (Wesson et al. 2003, Ercolano et al. 2003).  While the recombination line spectrum measured in A\,58 is not as well detected as that seen in A\,30, we detect recombination lines from C~{\sc ii} and O~{\sc ii} as well as the strong lines of He~{\sc i} and He~{\sc ii}.  We adopt a temperature of 500\,K to derive ORL abundances, based on the low values obtained from the O~{\sc ii} line ratios in Section~\ref{diagnostics}.

\subsubsection{Helium}
\label{Heabunds}

Abundances for helium were derived using atomic data from Smits (1996), accounting for the effects of collisional excitation using the formulae in Benjamin, Skillman \& Smits (1999).  Ionic and total helium abundances relative to hydrogen derived from helium recombination lines are given in Table~\ref{Hetable}.  Our adopted value of He$^+$/H$^+$ was derived from the $\lambda$4471, $\lambda$5876 and $\lambda$6678 lines, averaged with weights 1:3:1, roughly in proportion to their observed intensity ratios.  The helium abundances were calculated for a temperature of 500\,K.

\begin{table}
\centering
\caption{He/H abundance ratios, by number, for the central knot}
\label{Hetable}
\begin{tabular}{lll}
\hline
Species & Line & Abundance \\
\hline
He$^+$/H$^+$ & $\lambda$4471 & 3.05 \\
He$^+$/H$^+$ & $\lambda$5876 & 3.20 \\
He$^+$/H$^+$ & $\lambda$6678 & 2.99 \\
He$^+$/H$^+$ & Mean          & 3.13 \\
He$^2+$/H$^+$& $\lambda$4686 & 0.09 \\
He/H         &               & 3.21 \\
\hline
\end{tabular}
\end{table}

\subsubsection{Oxygen}

The emission from oxygen recombination lines is not as rich or as well detected as that found in Abell~30's knots.  Measurement of the oxygen lines is complicated by the fact that most of the lines we detect are significantly blended.  In the 4068--4076{\AA} region, oxygen recombination lines are blended with [S~{\sc ii}] CELs.  From the abundance derived for S$^{2+}$ from the $\lambda$6717/6731 lines, we calculated the expected strengths of the [S~{\sc ii}] lines at 4068 and 4076{\AA}, for a temperature of 15\,200\,K and the density of 2730\,cm$^{-3}$ measured from the $\lambda$6717/6731 ratio.  We find that [S~{\sc ii}] contributes 38\% of the total flux in this region, and we attribute the rest of the emission to O~{\sc ii}.

We clearly detect line emission at 4650{\AA}.  Because of the low excitation of the knot, we assume that no C~{\sc iii} emission is present.  By constraining line widths to be equal to that of the [O~{\sc iii}] line at 4959{\AA}, we obtained a reasonable fit to the broad feature using six gaussian curves, representing five lines from the O~{\sc ii} V1 multiplet, and [Fe~{\sc iii}] $\lambda$4658.  We derive O$^{2+}$/H$^+$ from the sum of the observed V1 multiplet lines, excluding the $\lambda$4669 line which is much too strong and may be affected by noise or blending.  We note that we do not detect the V1 multiplet line at 4642{\AA}.  According to Storey (2007), this line should have a strength about 60\% of that of the $\lambda$4649 line.  Its non-detection indicates that these lines, among the weakest measured in our spectra, have a low signal-to-noise ratio.  However, the abundances derived from the observed lines in the V1 multiplet are consistent with those derived from the V10 multiplet.

The lines we detect and the abundances derived from them are given in Table~\ref{heavyORLabunds}.  We derive abundances from individual lines and also from multiplets, by summing the fluxes of the observed components.  Our final O$^{2+}$/H$^+$ abundance of (0.46$\pm$0.10) by number is found by taking the average, with equal weights, of the abundances derived from the V1 and V10 multiplets and the 3d--4f transition at $\lambda$4089.

\subsubsection{Other elements}

We detect C~{\sc ii} lines at $\lambda$4267 and $\lambda$5342.  However, the abundance derived from the $\lambda$5342 line is much higher than the value derived from the $\lambda$4267 line.  It may be blended or misidentified and therefore we ignore it in our abundance determination.  We note that Ercolano et al. (2004) found that this line also yielded a carbon abundance much higher than other C~{\sc ii} lines in the high-adf nebula NGC~1501.

It is possible that C~{\sc iii} lines contribute to the observed flux at $\lambda$4647--52, but given the small amount of He$^{2+}$, we assume that the amount of C$^{3+}$ is negligible.

Doubts have long been raised about the reliability of C$^{2+}$ abundances derived from the $\lambda$4267 line, in light of the large discrepancy between them and abundances measured from UV CELs.  However, recent years have seen the detection in several nebulae of C~{\sc ii} lines from higher in the recombination cascade, whose observed intensities agree very well with the predictions of recombination theories (e.g. Tsamis et al. 2004).  It therefore seems very unlikely that another process besides recombination could be populating the upper level of the $\lambda$4267 transition.  We adopt C$^{2+}$/H$^+$ = 0.042, from the $\lambda$4267 value alone.

We do not detect any N~{\sc ii} or Ne~{\sc ii} recombination lines.

\begin{table}
\centering
\caption{Ionic abundances in the central knot of A\,58 derived from heavy element recombination lines}
\label{heavyORLabunds}
\begin{tabular}{llll}
$\lambda_{\rm 0}$ & Mult & $I_{obs}$ & C$^{2+}$/H$^{+}$ \\
\hline
4267.15 & V6    & 1.002 & 0.042 \\
5342.38 &       & 0.890 & 0.709$^a$\\
{\bf Adopted} & &       & {\bf 0.042} \\
\hline
$\lambda_{\rm 0}$ & Mult & $I_{obs}$ & O$^{2+}$/H$^{+}$ \\
\hline
4649.13 & V1    & 0.88 & 0.19 \\
4650.84 & V1    & 0.51 & 0.55 \\
4661.63 & V1    & 0.87 & 0.72 \\
        & {\bf V1} &   & {\bf 0.34} \\
4069.62 & V10   & 1.99 & 0.72 \\
4075.86 & V10   & 1.42 & 0.39 \\
        & {\bf V10} &  & {\bf 0.53}\\
4089.29 & V48a  & 0.94 & 0.51 \\
{\bf Adopted} & &      & {\bf 0.46 $\pm$ 0.10}\\
\hline
\end{tabular}
\begin{tabular}{l}
$^a$Not used in abundance determination due 
\\to possible blending or misidentification
\end{tabular}
\end{table}

\subsection{Total elemental abundances and abundance discrepancy factors}
\label{abundances}

To calculate total elements abundances, we adopt the ionisation correction scheme of Kingsburgh \& Barlow (1994).  The abundances derived are presented in Table~\ref{totalabunds}.  We also list the abundances derived in Abell~30 (Wesson et al. 2003).

For an element or ion X, the abundance discrepancy factor, adf(X), is defined as
\begin{equation}
adf(X) = \frac{X(ORL)/H}{X(CEL)/H}
\end{equation}

With the current data it is only possible to derive an adf for oxygen.  We find that adf(O$^{2+}$) and adf(O) are both 89.  These values are among the largest found in any ionised nebulae, larger than the values of 32 and 70 found for oxygen in the `normal' PNe NGC\,1501 (Ercolano et al. 2004) and Hf\,2--2 (Liu et al. 2006), but still considerably lower than the adfs of $\sim$700 seen in the polar knots of Abell~30 (Wesson et al. 2003).

\begin{table}
\centering
\caption{Total elemental abundances in the knot of A\,58 by number, on a logarithmic scale where N(H)=12.0, compared with those found in the knots of A\,30.}
\label{totalabunds}
\begin{tabular}{lllllll}
Object &       & He    & C     & N    & O     & Ne    \\
\hline
A58    & ORLs & 12.51 & 10.95 &  -    & 12.02 & -     \\
       & CELs & -     & -     &  9.21 & 10.03 &  9.92 \\
A30 J1 & ORLs & 13.03 & 11.65 & 11.49 & 12.15 & 11.51 \\
       & CELs & -     & -     &  8.88 &  9.26 &  9.70 \\
A30 J3 & ORLs & 13.07 & 11.66 & 11.43 & 12.10 & 11.99 \\
       & CELs & -     & 9.22  &  8.90 &  9.32 &  9.78 \\
\hline
\end{tabular}
\end{table}

\subsection{Comparison with previous results}

Abundances for A\,58's knot were previously derived by GM96.  They adopted a temperature of 12\,450\,K and $c({\rm H}\beta)$=0.29, and from CELs they derived value of O$^{2+}$/H$^+$ about eight times lower than our value and a value of O$^+$/H$^+$about three times higher.  They suggested that oxygen abundances were unreliable due to shock excitation of the knot.  However, we calculate the ratio of energy input from shocks and from photoionisation as follows:  the luminosity of the central star is 10$^4$L$_{\sun}$, and its effective temperature is 95\,000\,K (Clayton et al 2006).  Based on model atmospheres of hydrogen-deficient stars by Rauch et al. (2003), we estimate that 89\% of the energy from the star is emitted shortward of 912{\AA}, giving a total ionising luminosity of 8.9$\times$10$^3$L$_{\sun}$.  The impact of the stellar wind on the knot will be much more significant than the expansion of the knot into the low-density outer nebula in producing shocks.  Adopting \.M~=~-1.3$\times$10$^{-7}$\,M$_{\sun}$yr$^{-1}$ and v$_{\infty}$~=~2500\,km\,s$^{-1}$ from Clayton et al (2006), we calculate that the wind luminosity is only 67L$_{\sun}$, or 0.7\% of the ionising luminosity.  Therefore photoionisiation dominates over shock excitation in producing the observed spectrum.

GM96 found O$^{2+}$/O$^{+}$~=~0.05, while we obtain a value of $\sim$1.  This difference is partly accounted for by the lower extinction and temperature adopted by GM96: we derived abundances from their observed fluxes but using $c({\rm H}\beta)$=2.0 and $T_{\rm e}$=20\,800\,K, and obtained O$^{2+}$/O$^{+}$~=~$\sim$0.4.  The remaining factor of 2.5 may be due to the temperature of the central star, and thus the ionisation degree of the nebula, increasing during the nine years between GM96's observations and our own.

\section{Abundance patterns in the knot of A\,58 and other objects}

Table~\ref{comparisons} presents a comparison between the mass fractions of H, He, C, N, O and Ne that we have derived for the ejecta of A~58 and A~30, and those derived by other authors for the central stars of these two objects, as well as for a number of other objects, including PG~1159-035, which may be a descendant of a star similar to the nucleus of A~30; V4334~Sgr, which has been proposed to be a younger counterpart to A~58/V605~Aql; the shells of two old novae, DQ~Her and RR~Pic; and the ejecta of three neon novae. The nitrogen and neon abundances listed for the A~58 central knot are given in parentheses, since they are based on its measured CEL N:O:Ne ratios, normalised to its ORL oxygen abundance.

\subsection{Abell 30}

This analysis of the optical spectrum of A\,58 reveals many similarities with the much older knots of A\,30.  In both cases, extreme abundance discrepancy factors are observed, while ORL temperature diagnostics indicate the existence of very cold ionised plasma.  Modelling of Abell\,30 by Ercolano et al. (2003) demonstrated the physical plausibility of this situation, with the very high abundances of CNO coolants resulting in a very cold ionised core, with dust photoelectric heating generating the very high [O~{\sc iii}] temperatures in the outer part of the knot.  The C/O number ratio in both knots of A\,30 was found to be less than unity ($\sim$0.3 for both knots when derived from ORLs, and 0.8 for knot J3 from CELs), contrary to the predictions of the `born-again' scenario (Iben et al. 1983, Herwig 2001).  The knot of A\,58 is also found to be oxygen-rich, with a C/O ratio derived from ORLs of just 0.09.  The single-star VLTP model has serious difficulties in this scenario.

It is noticeable that the C/O mass fraction ratios measured for the A~58 and A~30 knots, 1/11 and 1/4 respectively, are much smaller than the ratios of of 8 and 2.7 measured for their central stars. In addition, the A~58 and A~30 knots have very large neon mass fractions. We conclude that the significant differences between the C, O and Ne mass fractions measured for the knots and for the central stars are real. Wesson et al. (2003) pointed out that the C/O $< 1$ ratios that they found for the knots of A~30 were not predicted by VLTP models for born-again nebulae. Werner \& Herwig (2006) have recently reviewed VLTP models, confirming that C/O ratios greater than unity are predicted, consistent with the central star C/O ratios but not the ejecta C/O ratios found for A~58 and A~30. In addition, VLTP models predict neon mass fractions $\leq$0.02, consistent with the values found for WCE central stars and PG~1159 stars, but much smaller than the neon mass fractions found in the knots of A~58 and A~30.

The A\,58 knot is presumed to have a red-shifted counterpart, on the opposite side of the star and completely obscured by dust.  The collimation of A\,30's knots argues against a single star being their source, and if A\,58 also has collimated outflows, as suggested by Pollacco et al. (1992), this could argue for some mechanism involving an accretion disc and polar jets in a double star system.  In contrast to A\,30, though, the knot of A\,58 is expanding quite rapidly.  While the knots of A\,30 are broadened by $<$20\,km\,s$^{-1}$ (Meaburn \& L\'opez 1996), the echelle spectra of Pollacco et al. (1992) yielded velocity FWHMs for the knot of 180\,km\,s$^{-1}$ from the [N~{\sc ii}] lines and 270\,km\,s$^{-1}$ from the [O~{\sc iii}] $\lambda$5007 line.

\subsection{Sakurai's Object}

Sakurai's Object (V4334 Sgr) was discovered in February 1996, and was initially thought to be a slow nova (Nakano et al. 1996).  However, its spectrum showed a rapid cooling, hydrogen abundance decline and an enhancement of s-process elements, over just a few months after its discovery, and it is thought to have undergone a final helium flash (Duerbeck \& Benetti 1996).

The spectrum of Sakurai's object at its maximum brightness is very similar to the spectrum of V605~Aql obtained at its maximum in 1919, with the appearance of a cool RCB star (Lundmark 1921).  Because of this, V605~Aql is often described as an older twin of Sakurai's Object.  The He:C:O mass ratios at the surface of Sakurai's object were found to be 85:5:3 in July 1996, compared to 54:40:5 for the central star of A\,58 (Clayton et al. 2006).  Our derived nebular abundances for A\,58's knot correspond to He:C:O mass ratios of 18:2:41.

Freshly ionised material has been detected around Sakurai's Object, indicating that a knot or knots similar to those seen in A\,58 may be forming.  The freshly ionised material is found to be expanding at 1500\,km\,s$^{-1}$ (Kerber et al. 2002).  At this expansion rate, its density will decline rapidly and it seems unlikely that any knot formed could survive to be seen as long after the event as the knots of A\,58 and A\,30.  The physical conditions and chemical abundances in the freshly ionised ejecta have yet to be determined.

\subsection{Classical novae}

Possible links between planetary nebulae showing high adfs and classical novae have been discussed by Wesson et al. (2003) and Liu et al. (2006).  In several cases, nova shells have been found to show very strong recombination line emission and evidence for very low temperatures.  One good example is the shell surrounding DQ Her, which was found by Williams et al. (1978) to have a Balmer jump temperature of $\sim$500K.  The C/O number ratio in the ejecta of DQ Her is 0.36 -- quite similar to that seen in the knots of Abell 30.  CP Pup shows similar spectral features with very strong recombination lines and a Balmer jump temperature of $\sim$800\,K, but in its case no carbon recombination lines are seen, implying a lower C/O ratio (Williams 1982).  

We have included in Table~\ref{comparisons} a summary of the elemental mass fractions measured for two old nova shells and for three `neon novae'. All five show C/O ratios of less than unity but the neon novae also show neon mass fractions that are comparable to the large values (0.1--0.4) that we have measured for the knots of A~58 and A~30. Models for neon novae (e.g. Starrfield et al. 1986; Politano et al. 1995) invoke the usual thermonuclear runaway on the surface of a white dwarf following mass transfer from a low mass companion, with the high neon abundances resulting from the fact that the runaway occurs on the surface of a high-mass (1.0-1.35~M$_\odot$) O-Ne-Mg white dwarf, some of whose subsurface material is mixed to the surface and ejected during the nova event. This suggests that the central stars of A~58 and A~30 might have high mass O-Ne-Mg cores, some of whose material may been brought to localised parts of the surface by the event that led to the ejection of the observed knots. If entropy barriers prevent such a scenario during a VLTP event then an alternative scenario could involve the transfer of mass from a companion on to localised parts of a massive white dwarf surface, leading to a thermonuclear runaway and the excavation and ejection of material from the O-Ne-Mg region of the white dwarf, whose thin surface layer still has C/O $>$ 1.

\begin{table*}
\centering
\caption{Comparison between the properties of Abell~58 and those of related objects.  Values in parentheses for A\,58 are based on CEL N/O and Ne/O ratios, scaled to the derived ORL oxygen abundance.}
\label{comparisons}
\begin{tabular}{lllllllll}
\hline
Name       & Type      &\multicolumn{6}{c}{Mass fractions}          & Reference \\
           &           & H     &He    & C     & N     &O    & Ne    &   \\
\hline
A 58       & knot      & 0.019 &0.250 & 0.021 &(0.043)&0.323&(0.345)& 1 \\
A 58       & WCE star  &       &0.54  & 0.40  &       &0.05 &       & 2 \\
A 30       & J1 knot   & 0.012 &0.519 & 0.065 & 0.052 &0.273& 0.078 & 3 \\
A 30       & J3 knot   & 0.010 &0.485 & 0.057 & 0.039 &0.208& 0.202 & 3 \\
A 30       & WCE-PG1159&       &0.41  & 0.40  & 0.04  &0.15 &       & 4 \\
PG1159-035 & PG1159    &$<$0.02&0.33  & 0.48  & 0.001 &0.17 & 0.02  & 5 \\
V4334 Sgr  & (RCrB)    & 0.003 &0.845:& 0.051 & 0.019 &0.027& 0.059 & 6 \\
DQ Her     & old nova  & 0.34  &0.095 & 0.045 & 0.23  &0.29 &       & 7 \\
DQ Her     & old nova  & 0.254 &0.123 & 0.046 & 0.302 &0.276&       & 8 \\
RR Pic     & old nova  & 0.53  &0.43  & 0.004 & 0.022 &0.006& 0.011 & 9 \\
V693 CrA   & neon nova & 0.408 &0.212 & 0.004 & 0.070 &0.069& 0.237 & 10\\
V4160 Sgr  & neon nova & 0.470 &0.338 & 0.006 & 0.058 &0.062& 0.065 & 11 \\
V1370 Aql  & neon nova & 0.053 &0.088 & 0.035 & 0.14  &0.051& 0.52  & 12 \\
\hline
\end{tabular}
\\{\small References: 1$)$ This paper; 2$)$ Clayton et al. (2006); 3$)$ Wesson et al. (2003); 4$)$Leunenhagen et al. (1993); 5$)$Jahn et al. 2007; 6$)$ Asplund et al. 1999 (July 1996 spectrum); 7$)$ Williams et al. (1978); 8$)$ from our re-analysis of the relative line intensities presented by Ferland et al. (1984); 9$)$ Williams \& Gallagher (1979); 10$)$ Vanlandingham et al. (1997); 11$)$ Schwarz et al. 2007; 12$)$ Snijders et al. 1987}
\end{table*}

\section{Discussion}

Of the five known hydrogen-deficient planetary nebulae (A\,30, A\,58, A\,78, IRAS 15154-5258 and IRAS 18333-2357), we have now carried out detailed ORL/CEL abundance studies of two of them.  In both cases, we find that the VLTP born-again scenario commonly invoked to account for the production of H-deficient material within an old planetary nebula, and which predicts C/O$>$1, cannot account for the abundance ratios observed, while the collimated outflows seen in both objects seem inconsistent with a single star at the centre.

The presence of knots of cold hydrogen-deficient material has commonly been invoked to account for the observed discrepancy in planetary nebulae whereby ORLs give much higher chemical abundances than CELs for heavy elements (e.g. Liu et al. 2000, Tsamis et al. 2004, Wesson et al. 2005).  The origin of this H-deficient material in normal nebulae is as yet unknown.  Given the uncertainty at the moment about how the knots in A\,30 and A\,58 have been produced, it is difficult to say if the proposed knots present in `normal' nebulae could have a similar origin.  However, we note that it is difficult for the evolution of a single star to explain the morphology and abundances in A\,30 and A\,58, and that recent work has suggested that most or all central stars of planetary nebulae could be binary systems (Moe and De Marco 2006).

However, spatially resolved spectroscopy and high resolution imaging of nebulae with high abundance discrepancy factors, such as NGC 6153, seem to suggest that knots in `normal' nebulae must be very small, $<$60\,AU across, numerous, and have a smooth distribution, as no clumps are seen in HST images, nor spikes in long slit spectra, but rather a smooth decline of adf from centre to edge, for example in NGC 6153 (Liu et al. 2000).  This is in contrast to the highly clumpy distribution of H-deficient material seen in Abell 30, Abell 58 and Abell 78.

\section{Acknowledgments}

This work is partly supported by a joint research grant co-sponsored by the Natural Science Foundation of China and the UK's Royal Society.  OD acknowledges funding from NSF-AST-0607111.  We thank the anonymous referee for a thorough and useful report.

\end{document}